\documentclass[prx,floatfix,twocolumn,showkeys,showpacs]{revtex4-1}
\usepackage{amsmath}
\usepackage{graphicx}
\usepackage{amsfonts}
\usepackage{amssymb}
\usepackage{bm}
\usepackage{epsfig,float,afterpage}
\usepackage[colorlinks,linkcolor=blue,citecolor=blue,urlcolor=blue]{hyperref}
\usepackage[usenames,dvipsnames]{color}
\usepackage{lipsum}
\def\beq{\begin{equation}}
\def\eeq{\end{equation}}
\def\bea{\begin{eqnarray}}
\def\eea{\end{eqnarray}}
\def\nn{\nonumber}
\def\ba{\begin{array}}
\def\ea{\end{array}}

\newcommand{\dg}{\dagger}
\def\one{1\hskip -1mm{\rm l}}
\newlength{\sizeonefig}
\newlength{\sizetwofig}
\begin{document}

\title{Electrical, thermal and thermoelectric transport in open long-range Kitaev chain }

\author{Averi Banerjee} 
%%\email{averi.banerjee@tict.edu.in}
\affiliation{Department of Basic Science and Humanities, Techno International Newtown, Kolkata-700156, India}
\author{Syeda Rafisa Rahaman}
\affiliation{Integrated Science Education \& Research Centre, Visva-Bharati University, Santiniketan-731235, India}
\author{Nilanjan Bondyopadhaya} 
\email{nilanjan.iserc@visva-bharati.ac.in}
\affiliation{Integrated Science Education \& Research Centre, Visva-Bharati University, Santiniketan-731235, India}

\begin{abstract}
We study electrical, thermal and thermoelectric transport in a hybrid device consisting of a long-range Kitaev chain coupled to two metallic leads at two ends. Electrical and thermal currents are calculated in this device under both voltage and thermal bias conditions. We find that the transport characteristics of the long-range Kitaev chain are distinguishably different from its short-range counterpart, which is well known for hosting zero energy Majorana edge modes under some specific range of values of the model parameters. The emergence of massive Dirac fermions, the absence of gap closing at the topological phase transition point and some special features of the energy spectrum which are unique to the long-range Kitaev chain, significantly alter electrical/thermal current vs. voltage/temperature bias characteristics in comparison with that of the short-range Kitaev chain. These novel transport characteristics of the long-range Kitaev model can be helpful in understanding nontrivial topological phases of the long-range Kitaev chain. 
\end{abstract}

\vspace{0.5cm}
%\date{\today}

%\pacs{: ~72.10.-d, ~05.60.-k, ~05.40.-a, ~73.50.Lw}
%\date{}
\maketitle

\begin{section}{Introduction}
Over the past decade, advancement in experimental physics has made it possible to realise topological superconductors in solid-state systems \cite{Alicea_2012,BeenakkerARCMP,Leijnse_2012,Sato_2017}.
%in the field of ultacold atoms trapped in optical lattice has opened up several possibilities to testify %theoretical studies of non-equilibrium transport in open quantum systems %\cite{BlochNP2012,Byrnes2020QuantumAO,HAFFNERPR2008,SafronovaRMP2018}. 
Among the proposed models of topological superconductors, the Kitaev chain has attracted much attention due to its unique ability to host topologically nontrivial zero energy Majorana Bound states (MBS), which are expected to be the critical ingredient to realise topological quantum computers \cite{ Kitaev_2001,Leijnse_2012,Sarma_npj2015}. Under appropriate conditions, a pair of MBSs is developed at the edges of a 1D Kitaev chain and these MBSs have significant ramifications on the non-equilibrium electrical \cite{Alicea_2012,Leijnse_2012,RoyPRB2012}, thermal, and thermoelectrical \cite{Banerjee2017,RamosPRB2016,FulgaPRB2011,Li_2017,SmirnovPRB2018,SmirnovPRB2019A,BhatDharPRB2020,BondyopadhayaJSP2022} transport through the Kitaev chain. It is to be noted that all the parameters, e.g. superconducting pairing terms and the hopping terms, of the Kitaev chain are short-ranged in the sense that these only couple electrons from the adjacent sites of the chain. Hence, this chain can also be referred to as the short-range Kitaev chain (SRK).

In recent years, an extended version, namely the long-range Kitaev chain (LRK), has become significant from both theoretical and experimental points of view. Long-range chain, which allows pairing and hopping terms to couple electrons from non-adjacent sites with strength decaying algebraically with the distance between the sites, shows stability against external perturbations \cite{Campa2014,Viyuela_PRB2016,Degottardi_PRB2013,Dutta_PRB2017,defenu2021longrange,SolfanelliJHEP2023}. Moreover, the long-range interactions induce novel correlations and entanglement behaviour \cite{FrancicaPRB2022}, delocalization of Majorana states in 1D LRK chain \cite{Alessandro_PRB}. On the experimental side, recent experiments using neutral atoms on the optical lattice coupled to photonic modes \citep{PerczelPRL2017,BettlesPRA2017}, using Shiba bound states induced by individual magnetic impurities on the surface of an s-wave superconductor \citep{PientkaPRB2013,PientkaPRB2014} and realization of programmable nonlocal interactions in an array of atomic ensembles within an optical cavity \cite{Pariwal2021} indicate the possibility of realizing long-ranged low-dimensional systems in a laboratory setup. Nowadays, many realistic quantum platforms featuring long-range power-law decaying interactions are gradually becoming an essential ingredient of modern quantum simulation \cite{defenu2021longrange}. The one-dimensional long-range  Kitaev chain represents a synergy between topological phases and long-range physics as topological properties of the Kitaev chain are not destroyed by the long-range interaction. Moreover, the study of LRK is very significant as it shows approximate behaviour of long-range Ising chain in a suitable limit \cite{Jaschke_2017} and also provides effective descriptions of periodically driven (Floquet) systems \cite{Benito_PRB}. 

In view of these, we find it interesting to explore the effect of long-range interaction on quantum transport in the LRK chain and compare these results with its short-range counterpart. Generally speaking, this long-range paring and hopping affect MBSs and the gap-closing phenomenon near the topological phase transition point (TPT). Zero energy MBSs of the standard (short-range) Kitaev chain transformed into massive Dirac fermions when the long-range interaction is switched on as this causes the wavefunctions of the two boundary Majorana modes to hybridize by overlapping \cite{VodolaNJP2016,Patrick_PRL2017,GiulianoPRB2018,Viyuela_PRB2016,Alessandro_PRB}. Moreover, the long-range interaction affects the spatial extent of the eigenfunctions of the LRK chain, and also modifies the density of states. Specifically, the bulk states of LRK become more delocalized as compared to that of the SRK, and these delocalized states generally assist the transport process \cite{Abumwis_PhysRevLett2020}. 

We consider an N-TS-N junction where the middle topological superconductor (TS) is an LRK/SRK chain, and the leads/baths are made of normal metal (N). It is known that the study of electrical and thermal currents in the SRK chain has been discussed in several literature, but such transport in the LRK chain has yet to be widely studied. To address this issue, we use the well-known quantum Langevin equations \& Green’s function method (LEGF) \cite{DharSenPRB2006,RoyPRB2012,BondyopadhayaJSP2022,BhatDharPRB2020} to study electrical, thermal, and thermoelectric currents in the LRK chain. Baths are considered to be at thermal equilibrium described by Fermi distribution functions, which depend on the temperature and chemical potential of the corresponding baths. Voltage or temperature bias is created between the baths, which drive electrical/thermoelectrical and thermal current through the N-TS-N junction. We study electrical/thermal current variation as a function of voltage/thermal bias for both LRK and SRK chains and compare their relative behaviours. There are some interesting studies of the Fano factor related to the electrical current, which is the ratio of shot noise and the electrical current, as a function of voltage bias in Ref.  \cite{GiulianoPRB2018}, which emphasizes the contrasting behaviour of the same for LRK and SRK chains. We find that the study of electrical, thermal, and thermoelectrical currents as a function of different types of biases (voltage/temperature) is also interesting for LRK chain as these observables are greatly influenced by the delocalized states, the modified density of states, the finite mass of subgap edge states, and the absence of gap closure at TPT which are categorically different from the characteristics of short-range Kitaev chain \cite{Viyuela_PRB2016}. Hence the current characteristic at TPT point can be a good indicator of formation of massive Dirac edge mode \cite{Viyuela_PRB2016}. Moreover, in thermoelectric transport, we observe a change of direction/modulation of current with the variation of
%in the variation of thermal current w.r.t. voltage bias, 
the chain parameters, e.g. on-site energy. This type of behaviour of thermoelectric current is also observed in different topological chains like the Su-Schrieffer-Heeger(SSH) and Rice-Mele (RM) chains\cite{LimaSR2023}.

%The thermopower is positive or negative depending whether the charge carriers are holes or electrons,respectively.

The article is organized as follows. In Sec. \ref{sec:chain}, we discuss the Hamiltonian of the Kitaev chain with long-range pairing and long-range hopping terms. The general expressions for electrical and thermal currents are discussed in Sec. \ref{sec:currents}. In Sec. \ref{sec:results} comparative behaviour of electrical, thermal and thermo-electrical currents are studied for both LRK and SRK chains. We conclude with a short discussion in Sec. \ref{sec:conc}.
\end{section}

\begin{section}{Model Hamiltonian}
\label{sec:chain}

We consider a 1D long-range Kitaev (LRK) chain in which hopping terms contain both long-range interaction and phase factor whereas real pairing terms contain only long range interaction. The tight-binding Hamiltonian for this chain in open system geometry is given by \cite{VodolaPRL2014,VodolaNJP2016,AleccePRB2017,Viyuela_PRB2016},
% (we choose $\hbar=1$)
\bea
\frac{\mathcal{H}}{\hbar}&=&-\sum_{l=1}^{N} \sum_{r=1}^{N-l} \left( \gamma_r\,   c^{\dg}_{l}c_{l+r} + \Delta_r\, c^{\dg}_{l}c^{\dg}_{l+r} + \text{H.c.} \right) -{\epsilon} \sum_{l=1}^{N}(2c^{\dg}_{l}c_{l}-1) \, ,\nn \\
\label{HLRK}
\eea
where $c_l$ is the fermionic annihilation operator at the $l$-th site, $N$ is the number of lattice sites of the chain, and $\epsilon$ is the on-site energy. Hopping and superconducting pairing terms which encode long-range interactions between $l$-th and $(l+r)$-th sites, read $\gamma_r = \gamma_0  r^{-\eta}$ and $\Delta_r = \Delta_0 r^{-\alpha}$, respectively, and we choose both as real in this study. The parameters $\eta $ and $\alpha$ control the power-law decay of the hopping and pairing terms. 

The power law decay behaviour of hopping and superconducting pairing terms is determined by the exponents $\eta \, (\eta >0 ) $ and $\alpha \, (\alpha >0) $, respectively. This chain reduces to a standard short-range Kitaev (SRK) chain in the limit $\eta \rightarrow \infty$ and $\alpha \rightarrow \infty$. On the other hand, fermionic edge modes of LRK become massive for $\eta <1$ and $\alpha <1$, due to the non-zero overlap of the Majorana edge modes. Thus, different regimes of power exponents reveal exciting features of the LRK chain compared to the SRK chain. 

In order to study non-equilibrium steady-state transport in this LRK chain, we couple this chain with two metallic baths at two ends to construct a metal-topological superconductor-metal (N-TS-N) hybrid device. Baths are kept at thermal equilibrium before they are connected with the LRK chain at time $t_0$; after that the whole device evolves with time ($t > t_0$) to reach non-equilibrium steady-state. In non-equilibrium steady-state, transport properties can be derived by first taking the limit $t_0 \rightarrow -\infty$, and then integrating out the bath operators using quantum Langevin equations and Green’s function (LEGF) method, which has been extensively discussed in many previous articles \cite{DharSenPRB2006,Roy_Dhar_PRB2007,RoyPRB2012,RoyBondyopadhayaPRB2013,BhatDharPRB2020,BondyopadhayaJSP2022}.   
For the sake of completeness, we have added a brief discussion about LEGF method in Appendix~\ref{App1}

The metallic baths are modelled by semi-infinite free-electron tight-binding chains. First site ($l=1$) of the LRK chain is coupled to the left bath, which is kept at the chemical potential $\mu_L$, and temperature $T_L$ while the last site ($l=N$) of the same is coupled to the right bath which is kept at chemical potential $\mu_R$, and temperature $T_R$. Left and right metallic bath Hamiltonians $\mathcal{H}_M^p$ ($p=L,R$), and the corresponding tunnel couplings between the baths and LRK chain, $\mathcal{H}_T^p$ are given by 
\bea
\mathcal{H}_M^p &=& - \gamma_p \hbar \sum_{\beta=1}^{\infty} \left(c^{p \dg}_\beta \, c^p_{\beta+1}+c^{p \dg}_{\beta+1} \, c^p_{\beta} \right) \, ,
\label{HBath}\\
\mathcal{H}_T^p &=& - \gamma_p ' \hbar \, \left( c^{p \dg}_1 \, c_{l_p}+ c^\dg_{l_p} c^p_1 \right) \, ,
\label{HTun}
\eea
where $l_L=1$ and $l_R=N$ for an LRK chain with $N$ sites. Here, $\gamma_p$ represents hopping strength of the metallic bath and $\gamma_p '$ controls the strength of the tunnel coupling between the normal baths and the LRK chain. Clearly $c^{p \dg}_\beta$ represents an electron creation operator on the $\beta$-th site of the $p$-th bath. The initial equilibrium correlations for the baths are given by
\beq
\langle c^{p \dg}_\beta (t_0)c^p_{\beta '} (t_0) \rangle =\sum_k \psi^{p \,*}_k(\beta)\psi^{p }_k(\beta ') f( \omega_k^p,\mu_p,T_p), \nn
\label{ini}
\eeq
where $\sum_{\beta '} (\mathcal{H}_M^p)_{\beta \beta '}\, \psi^{p }_k(\beta ')=\omega_k^p \,\psi^{p }_k(\beta)$ and, $f(\omega,\mu,T)=(\exp\{(\omega -\mu)/k_B T \}+1)^{-1}$ denotes the Fermi function \cite{DharSenPRB2006}. By definition, $\psi^{p }_k(\beta ')$ is the $\beta '$-th component of the $k$-th eigenfunction of $p$-th metallic bath Hamiltonian. For example, if $\Psi^{L}_k$ represents the $k$-th eigenfunction of left metallic bath Hamiltonian ($\mathcal{H}_M^L$), it can be expressed in the form of a tuple $\Psi^{L}_k \equiv \{\psi^{L}_k(1), \psi^{L}_k(2), \dots, \psi^{L}_k(\beta '),\dots \}$.

\end{section}

\begin{section}{Electrical and thermal currents}
\label{sec:currents}

We apply the LEGF approach to study electrical, thermal and thermoelectrical transport in LRK chain under steady-state conditions \cite{DharSenPRB2006,DharJSP2006,RoyPRB2012,RoyBondyopadhayaPRB2013,BhatDharPRB2020,BondyopadhayaJSP2022}. It is well known that the LEGF approach provides a more direct method to study transport in an open system framework while treating both the wire and the baths in microscopic detail. Since the Hamiltonians of the tunnel junctions $\mathcal{H}_T^p$ (\ref{HTun})  do not contain any long-range interaction terms, the expression for both electrical current and thermal current remain same as that of the short-range Kitaev chain, which are explicitly calculated in Ref. \cite{DharSenPRB2006,BondyopadhayaJSP2022}. 

Since the tunnel coupling Hamiltonians are metallic, the conservation of electrical charges holds across the tunnel junctions. Hence, the continuity equation for electrical charges helps to define the electrical current across the left and right junctions, $J_{\rm L}^e$ and $J_{\rm R}^e$ respectively :
\bea
J_{\rm L}^e &=&i e \, \gamma'_{L} \langle (c_{1}^{\dg} (t)c_{1}^L(t)-c_{1}^{L \dg}(t) c_{1}(t))\rangle 
~~~~~~~~~~~~~~\nn \\
&=&  -2 e \gamma'_{L} \text{Im} [\langle c_{1}^{\dg} (t)c_{1}^L (t) \rangle ] \, , \nn \\
J_{\rm R}^e &=&i e \, \gamma'_{R} \langle (c_{1}^{R\, \dg} (t)c_{N}(t)-c_{N}^{ \dg}(t) c_{1}^{R}(t))\rangle 
~~~~~~~~~~~~~~\nn \\
&=& -2 e \gamma'_{R} \text{Im} [\langle c_{1}^{ R \, \dg} (t)c_{N} (t) \rangle ] \, , \label{elec_c}
\eea
where $e$ is the magnitude of the electronic charge. The expectation $\langle .. \rangle$ denotes averaging over the initial density matrix of baths. The initial density matrices of baths are calculated at $t=t_0$, when left and right baths are isolated and described by grand canonical ensembles at temperature and chemical potential given by $(T_L,\mu_L)$ and $(T_R,\mu_R)$, respectively. Since the number of degrees of freedom of a bath/reservoir is much greater than that of the middle wire, the initial density matrix of a bath does not undergo any perceptible changes even after $t>t_0$ when the tunnel junction is established between the bath and the wire.

 According to our convention, $J_L^e$ ($J_R^e$) represents the electrical current from the left bath to the wire (wire to the right bath). Since thermoelectric current is electrical current generated due to the temperature difference between two baths, these expressions of  $J_{\rm L/R}^e$ are also used in \ref{sec:thermoelectrical}.

Using the continuity equation for conserved energy across the junctions between LRK chain and the baths, we define left junction energy/thermal current ($J^u_{\rm L}$) and right junction energy/thermal current ($J^u_{\rm R}$) as :
\bea
J^u_{\rm L} &=& 2 \hbar \gamma_L' \left( \gamma_0 \text{Im}[\langle c_2^{\dg} (t)c_{1}^L (t) \rangle ]+\Delta_0\text{Im}[\langle c_2^{\dg} (t)c_{1}^{L \,\dg} (t) \rangle ] \right) \nn \\
&& +2 \hbar  \gamma_L' \epsilon \, \text{Im}[\langle c_1^{\dg} (t)c_{1}^L (t) \rangle ] \, , \nn \\
J^u_{\rm R} &=& 2 \hbar \gamma_R' \left( \gamma_0 \text{Im}[\langle c_{1}^{ R \dg} (t)c_{N-1}(t) \rangle ]-
\Delta_0\text{Im}[\langle c_1^{ R \,  \dg} (t)c_{N-1}^\dg(t) \rangle ] \right) \nn \\
&& +2 \hbar \gamma_R' \epsilon \, \text{Im}[\langle c_1^{R\, \dg} (t)c_{N} (t) \rangle ]  \, .\nn \\
\label{energy_c}
\eea
 
Non-equilibrium steady-state transport is realised by keeping two fermionic baths at biased conditions. We study voltage and thermal bias separately. Under voltage biasing, chemical potentials of baths are kept symmetrically, i.e. $\mu_L=-\mu_R= eV/2$ and $\mu_L-\mu_R = eV$; in the case of temperature bias, the bath's temperatures are kept asymmetrically, i.e. $T_L=T+ \Delta T, \, T_R=T$ ($\Delta T>0$). Under such voltage and thermal bias, we explore three main transport properties, namely (i) electrical current ($J^e$) as a function of the voltage difference ($eV$), (ii) thermoelectric current ($J^e$) as a function of temperature difference ($\Delta T$), and (iii) thermal current ($J^u$) as a function of temperature difference ($\Delta T$). In all studies in the subsequent sections, we choose hopping strengths equal for two baths, i.e. $\gamma_L=\gamma_R=\gamma_p$, and these are connected with the LRK chain by identical contacts ($\gamma_L '=\gamma_R '=\gamma_p'$).  For simplicity, we neglect any influence of voltage/temperature bias on the superconducting pairing term $\Delta_0$ because the essential qualitative features of the steady-state transport in such a device do not depend on the exact value of $\Delta_0$ \cite{Lobos_NJP2015}.

\end{section}

\begin{section}{Result and Discussions}
\label{sec:results}
The transport characteristics of the SRK chain show several unique features. Some of these are perceived as the  most useful tools to detect topological phase transitions and the presence of Majorana fermions in nontrivial topological phase \cite{BeenakkerARCMP,Alicea_2012,Leijnse_2012,Sau_PRB2010,TewariPRL2008,RoyPRB2012}. Here, we find that the transport characteristics of the LRK chain also show some distinct features which are quite different from its short-range variant. Throughout the rest of the section, we choose $\alpha=0.5$ and $\eta=0.5$, in which the long-range nature of interaction dominates. However, $\alpha=10.0$ and $\eta=10.0 $ are used to study the behaviour of its short-range counterpart. It is to be noted that only for electrical/thermal current vs voltage characteristics, we choose $N=20$, and for the rest of cases, $N=15$ are chosen. Moreover, for notational convenience, $e=1$, $k_B=1$ and $\hbar=1$ are chosen for the rest of the article.

\begin{subsection}{Electrical current in the LRK chain}
\label{sec:electrical}
 
In this subsection, we perform a comparative study of electrical current as a function of bias voltage for LRK and SRK chains in (i) $\epsilon=0$ and (ii) $\epsilon= 2 \gamma_0$ limits.  For both cases, right and left baths are kept at same temperature $T_L=T_R=0.02$. In Fig. \ref{elec_V_0}, corresponding to ${\epsilon} = 0$, electrical currents increase similarly for both LRK and SRK chains until $V=0.2$. Then currents in both models saturate at the same value due to the presence of the same energy gap in their respective energy spectrums. This sharp increase occurs due to low-energy MBSs in both chains. As the voltage increases, bulk states above the band gap participate in the transport. Due to the long-range interaction, these bulk states in the LRK chain are more delocalized and, thus, abet transport in the LRK chain. Hence, current increases at a faster rate in the LRK chain. However, this current saturates at a relatively lower value of applied voltage bias when the applied voltage bias surpasses the bandwidth of the wire. It should be noted that the maximum value of the current strongly depends on the number of quasiparticle modes of the superconductor that can be excited with the given voltage bias. For the set of parameters used in Fig. \ref{elec_V_0}, numerical diagonalization of LRK Hamiltonian reveals that out of total $40$ quasiparticle states, there are four quasiparticle states with energies: $4.61995,-4.61995,1.41351,-1.41351 $. Since the maximum biasing voltage used in Fig. \ref{elec_V_0} i.e. $V=2.5$ can excite states between $(-1.25,1.25 )$, these four quasiparticle states are not excited by $V=2.5$, and they do not take part in electrical transport through the LRK chain. Clearly, $V \sim 2.0 $ can not excite all the quasiparticle states in the LRK chain. However, for the SRK chain all $40$ energy states lie in between $(-0.990833,0.990833)$; thus all these states can be excited with $V \sim 2.0$. Since, for these values of the parameters, $V \sim 2.0 $ can excite more number of quasiparticle states in the SRK chain compared to LRK, the maximum value of the current is higher in the SRK chain, as depicted in Fig. \ref{elec_V_0}.

The situation is more interesting for $\epsilon= 2 \gamma_0$ when gap closure occurs in the short-range chain. However, due to the long-range interaction in the LRK chain, the energy spectrum always remains gapped. This $\epsilon= 2 \gamma_0$  point is also known as the topological phase transition (TPT) point for the SRK chain. In this phase, owing to the symmetrically placed positive and negative energy branches, the band gap for the LRK chain becomes twice the energy of the lowest energy state of the upper branch ($E_{min}^+$). Interestingly, we find that the bandgap ($\Delta E= 2E_{min}^+$) becomes independent of the length of the chain, i.e. $N$, but it depends on other system parameters, namely $\gamma_r$, $\Delta_r$ and $\mu$. Electrical current vs voltage at TPT point is plotted in Fig. \ref{elec_V_1}. For the short-range chain, electrical current increases linearly with voltage, which is expected as the energy spectrum of the SRK chain is gapless in this phase. In the presence of long-range interaction current remains zero till the $V=2 E_{min}^+$ (for this specific case, $E_{min}^+=0.396$); after that it starts increasing at a faster rate due to two reasons. Firstly, delocalized bulk states have higher quasiparticle density ($| \psi(l)|^2$) at the edges as well as in the bulk; therefore more quasiparticles can flow through the junctions via these bulk channels. Secondly, long-range interaction modifies the energy spectrum so that more bulk states are bunched up just above and below the energy gap. In the SRK chain bulk states are uniformly distributed over the entire energy range, whereas in the LRK chain, bulk states are densely distributed just above and below the band gap region but sparsely distributed in the upper part (lower part) of the upper band (lower band). Thus, a slight increase in voltage above the $\Delta E$, excites a large number of bulk states in the LRK chain compared to that of the SRK chain. Fig. \ref{bunch}, clearly depicts the bunching/clustering of energy levels of the LRK chain and its comparison with almost uniformly distributed energy levels of SRK chain.

These above-mentioned properties of the energy spectrum of LRK help the current to increase at a faster rate and eventually overtake that of the SRK chain at a particular value of $V$. For example, in Fig. \ref{elec_V_1}, this overtaking occurs at $V_0 \sim 1.0 $ as the numerical diagonalization of the Hamiltonian shows that majority of the delocalized bulk states are bunched up within the energy range $\approx |V_0-E_{min}^+|$. For $N=20$ (all other parameters are the same as Fig. \ref{elec_V_1}), $32$ eigenstates out of a total $40$ eigenstates are distributed symmetrically about the zero energy within the energy range $|V_0-E_{min}^+|$ with $E_{min}^+=0.369$ (See Fig. \ref{bunch}(b)).  As it appears from Fig. \ref{bunch}, there are more number of quasiparticle states lie between $(-1.25,1.25)$ for LRK chain in contrast to SRK chain, hence the maximum value of electrical current is higher in LRK chain than SRK chain for the bias voltage $V =2.5$ which can excite states between $(-1.25,1.25)$. This feature is depicted in Fig. \ref{elec_V_1}.

It is worth noting that the value of $E_{min}^+$ for the LRK chain does not vary with the length ($N$) of the chain for the region $\epsilon = 2\gamma_0$, which corresponds to a TPT point. Therefore the bandgap $\Delta E=2 E_{min}^+$ does not scale with $N$ ($ \Delta E \sim N^0$) \cite{Viyuela_PRB2016}. Hence, this characteristic behaviour of $J^e$ at $\epsilon = 2\gamma_0$ is also expected to be observed in the LRK chains with larger lengths and even in the thermodynamic limit. Moreover, this qualitative nature of the clustering of eigenstates as mentioned in the preceding paragraph, remains the same even if we increase the $N$ value. For example, $160$ eigenstates out of $200$ eigenstates of a LRK chain having length, $N=100$ (other parameters are same as Fig. \ref{elec_V_1}), are localized within a bandwidth  $|V_0-E_{min}^+|$ just above and below the bandgap, while  $E_{min}^+=0.365$ and $|V_0| \sim 1.1$, 
Therefore, we can infer that the clustering behaviour of eigenstates of the LRK chain is a characteristic feature of the long-range interactions.

\begin{figure}[htb]
\centering{\includegraphics[width=1\linewidth]{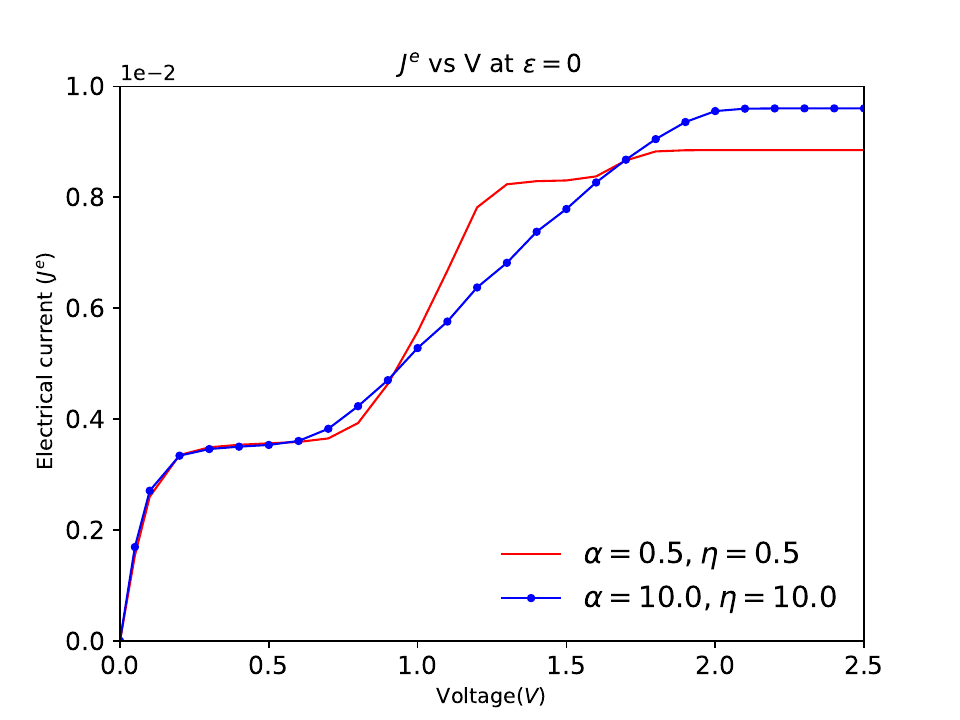}}
%\centering{\includegraphics[width=1\textwidth]{Elec_cur_vs_V_epsi_0_gamma_05.pdf}}
\caption{Plot of electrical current ${J^e}$ vs ${V}$, ${{N}=20,|{\epsilon}|=0.0,|{\gamma_0}|=0.5,{\Delta}=0.15},\gamma_{p} =1.0 ,\gamma'_{p}=0.1 $}
\label{elec_V_0}
\end{figure}

\begin{figure}[htb]
%\centering{\includegraphics[width=1\textwidth]{Elec_cur_vs_V_epsi_1_gamma_05.pdf}}
\centering{\includegraphics[width=1\linewidth]{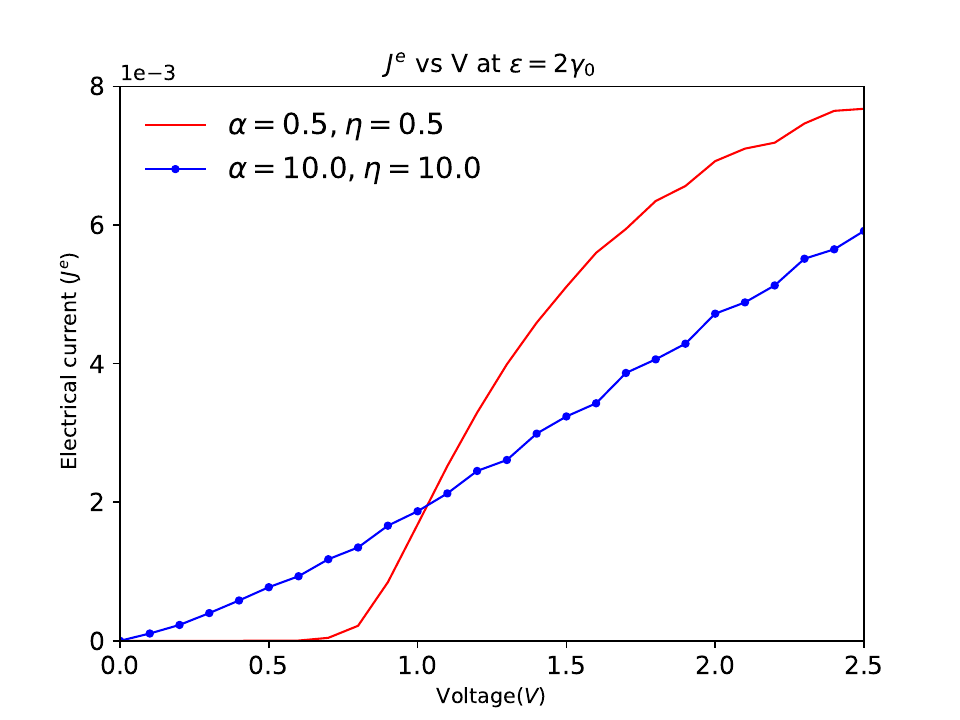}}
\caption{Plot of electrical current ${J^e}$ vs ${V}$, ${{N}=20,|{\epsilon}|=1.0,|{\gamma_0}|=0.5,{\Delta}=0.15},\gamma_{p} =1.0 ,\gamma'_{p}=0.1  $}
\label{elec_V_1}
\end{figure}

\begin{figure}[htb]
%\centering{\includegraphics[width=1\textwidth]{Elec_cur_vs_V_epsi_1_gamma_05.pdf}}
\centering{\includegraphics[width=1\linewidth]{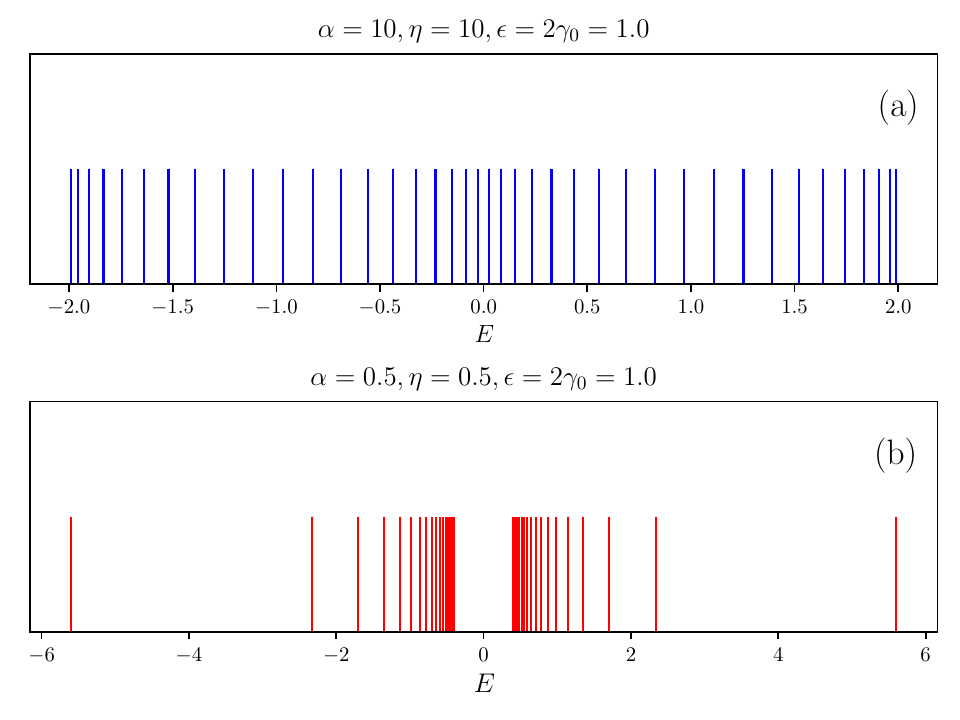}}
\caption{Graphical representation of energy levels' distributions of SRK chain and LRK chain at TPT point. Positions of energy levels are represented by the red/blue spikes. Each spike signifies presence of one eigenstate with an energy as determined from the location of the spike along the $E$ axis. Upper (Lower) panel represents energy levels of SRK (LRK) chain. In both panels, ${{N}=20,|{\epsilon}|=1.0,|{\gamma_0}|=0.5,{\Delta}=0.15}$. (a) Almost uniformly distributed energy levels in SRK chain ($\alpha=\eta=10$), and (b) Bunching of energy levels and the absence of gap closing at TPT point in LRK chain ($\alpha=\eta=0.5$) }
\label{bunch}
\end{figure}

\end{subsection}

\begin{subsection}{Thermoelectric current in the LRK chain}
\label{sec:thermoelectrical}
Variation of thermoelectric current with temperature bias reveals some interesting properties of LRK chain. 
In Fig. \ref{elec_dT_all}, we plot thermoelectric currents at the left junction as a function of the temperature 
difference between the left and right bath ($\Delta T =T_L-T_R $) for four different values of $\epsilon$. In all 
these plots, $T_R$ is kept at $0.02$ while $T_L$ is varied from $0.02$ to $0.22$ ($\Delta T \in \{ 0.0,\dots,0.2\}
$). Fig. \ref{elec_dT_all}(a) shows that the thermoelectric current saturates at a higher value in the case of LRK chain compared to that of SRK chain for ${\epsilon} = 0.0$. We think this rapid increase is owing to the delocalized subgap and bulk states of the LRK chain. In Fig. \ref{elec_dT_all}(b),  at the TPT point ($\epsilon=2 \gamma_0$), the current in LRK shows a higher growth rate after remaining close to zero initially. As $\Delta T$ increases enough to excite densely spaced bulk states above the band gap, the current shows rapid growth in the LRK chain. It is very similar to the phenomenon that we already observed for $J^e$ in Fig. \ref{elec_V_1}. In Fig. \ref{elec_dT_all}(c), the polarity of thermoelectrical currents for LRK and SRK chains are opposite to each other. Fig. \ref{elec_dT_all}(d) shows that $J^e$ in the LRK chain initially increases with a positive gradient, and around $\Delta T=0.1$ it reaches the maxima, then it decreases gradually and eventually becomes negative. However, for this set of parameters, $J^e$ in the SRK  chain does not make such a positive to negative transition, but grows gradually in the negative direction. This marks a significant difference between the LRK and SRK chains in the field of thermoelectrical transport. We think this change of
polarity happens due to the alteration of the nature of the majority carrier with respect to the variations of on-site energy ($\epsilon$) and junction temperature ($\Delta T$). As $\epsilon$ and $\Delta T$ vary, the majority carriers toggle between electron-like quasiparticles and hole-like quasiparticles and cause the $J^e$ to change its polarity. Moreover, we find interesting positive to negative transition of thermoelectric current for intermediate values of $\epsilon$. This behaviour can be attributed to the fact that the polarity of the thermoelectric current depends on whether electron-like or hole-like quasiparticles play the role of the majority carrier in such transport.

\begin{figure}[!htb]
\centering{\includegraphics[width=1\linewidth]{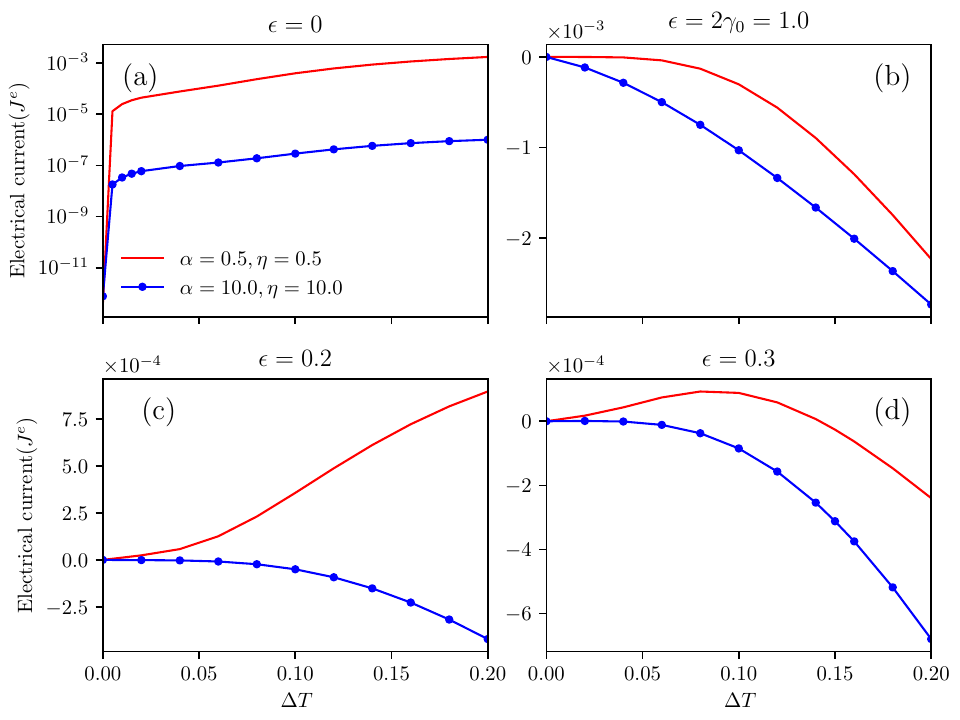}}
\caption{Plot of electrical currents ${J^e}$ vs $\Delta T$, $N=15,|\gamma_0|=0.5,\Delta=0.15,\gamma_{p} =1.0 , \gamma'_{p} =0.25 $, (a)$\epsilon=0$, (b)$\epsilon=1$, (c)$\epsilon=0.2$, and (d)$\epsilon=0.3$}
\label{elec_dT_all}
\end{figure}

\end{subsection}

\begin{subsection}{Thermal current in the LRK chain}
\label{sec:thermal}

We study the variation of thermal current as a function of  ${\Delta}T$ in Fig. \ref{en_dT_V0} and Fig. \ref{en_dT_V1} for different values of on-site energy. In Fig. \ref{en_dT_V0}, which corresponds to $\epsilon=0$, thermal current increases more rapidly for LRK than for the SRK chain.
% Since the low energy mid gap state in LRK chain is strongly delocalized than MBS of SRK chain, it 
Delocalization of subgap states and bunching up of bulk states help the thermal current in the LRK chain to grow much faster, as depicted in the figure. 

Since $\epsilon=2 \gamma_0=1.0$, i.e. TPT point, corresponds to the gap-closing phase of the SRK chain, the thermal current in the SRK chain starts increasing right from zero as $\Delta T$ increases, and this behaviour is depicted in Fig. \ref{en_dT_V1}. However, in the case of the LRK chain, energy gap is not closed for $\epsilon=2 \gamma_0=1.0$. So, we observe that the $J^u$ remains almost zero till $\Delta T \sim 0.06$, then it starts increasing, and around $\Delta T = 0.15$ it overtakes that of the SRK chain owing to the excitations of bunched-up bulk states situated above the bandgap.

\begin{figure}[!htb]
\centering{\includegraphics[width=1\linewidth]{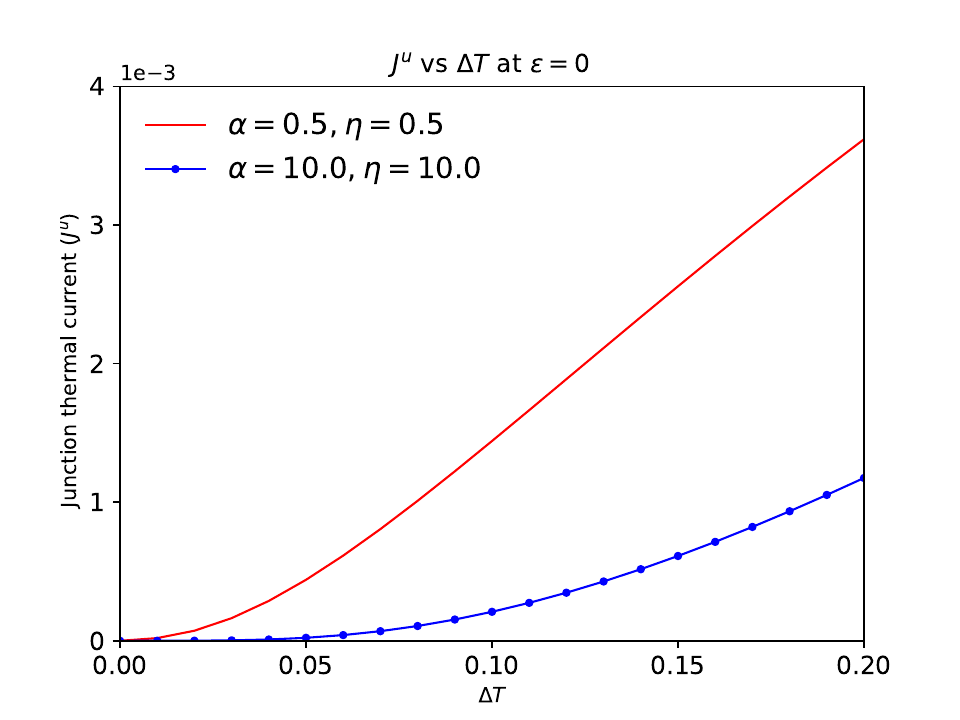}}
\caption{Plot of thermal currents ${J^u}$ vs $\Delta T$, $N=15, |{\epsilon}|=0.0, |\gamma_0|=0.5,\Delta=0.15,\gamma_{p} =1.0 , \gamma'_{p} =0.25 $}
\label{en_dT_V0}
\end{figure}

\begin{figure}[!htb]
\centering{\includegraphics[width=1\linewidth]{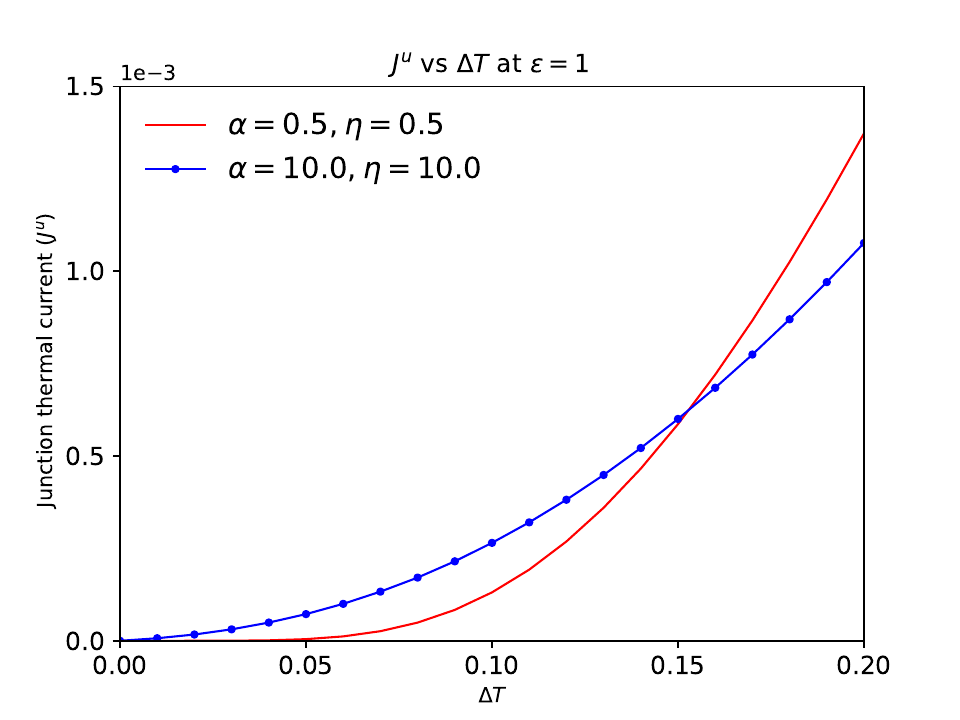}}
\caption{Plot of thermal currents ${J^u}$ vs $\Delta T$, $N=15, |{\epsilon}|=1.0, |\gamma_0|=0.5,\Delta=0.15,\gamma_{p} =1.0 , \gamma'_{p} =0.25 $}
\label{en_dT_V1}
\end{figure}

Further, we study the variation of $J^u$ with bias voltage, $V$. Fig. \ref{en_V_V0} shows a variation of $J^u$ when $\epsilon=0$. Both plots show the initial increase of the thermal current in the negative direction as  voltage increases, but then its absolute value decreases again. Interestingly, in  Fig. \ref{en_V_V0}(b), $J^u$ in the SRK chain changes its polarity around $V\sim 1.1$ and increases in the positive direction and finally saturates. On the other hand, in  Fig. \ref{en_V_V0}(a) $J^u$ in the LRK chain does not change polarity, but it shows oscillating behaviour with the variation of $V$ before showing saturation beyond $V \sim 2$. The magnitude of $J^u$ is much larger in the case of LRK due to delocalized subgap states and the bunched-up bulk states. We interpret this oscillating behaviour of $J^u$ with the variation of  $V$ as a ramification of the change of relative contributions of electron-type and hole-type quasiparticles to the transport. It is evident from the plot that this relative contribution also varies with the change of on-site energy and voltage bias.

In Fig. \ref{en_V_V1}, we plot the thermal currents at the TPT point. At $\epsilon=2 \gamma_0=1.0$, magnitude of $J^u$ in the SRK chain gradually increases in the initial part of the voltage range, whereas that for LRK chain remains close to zero due to the gap in their energy spectrum. However, above $V \sim 1.0$, voltage bias becomes sufficient to excite delocalized bulk states of LRK, which, in turn causes $|J^u|$ to increase rapidly in the LRK chain than its short-range counterpart.

\begin{figure}[!htb]
\centering{\includegraphics[width=1\linewidth]{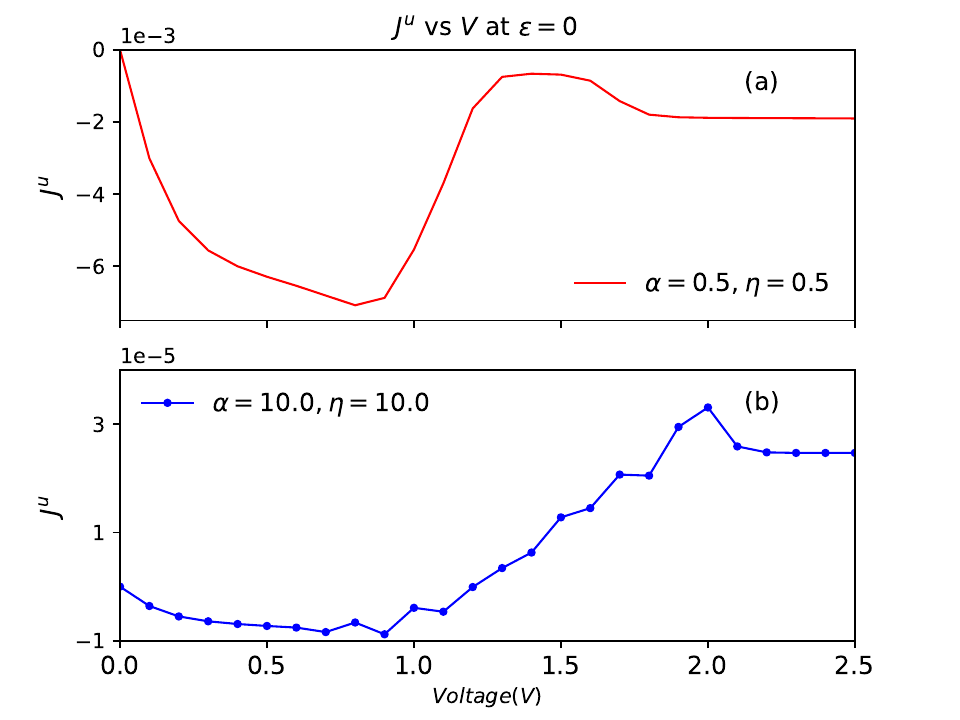}}
\caption{Plot of thermal currents ${J^u}$ vs $V$, $N=20, |{\epsilon}|=0.0, |\gamma_0|=0.5,\Delta=0.15,\gamma_{p} =1.0 , \gamma'_{p} =0.25 $. (a) Upper panel represents ${J^u}$ vs $V$ in LRK chain, (b) Lower panel represents ${J^u}$ vs $V$ in SRK chain.}
\label{en_V_V0}
\end{figure}

\begin{figure}[!htb]
\centering{\includegraphics[width=1\linewidth]{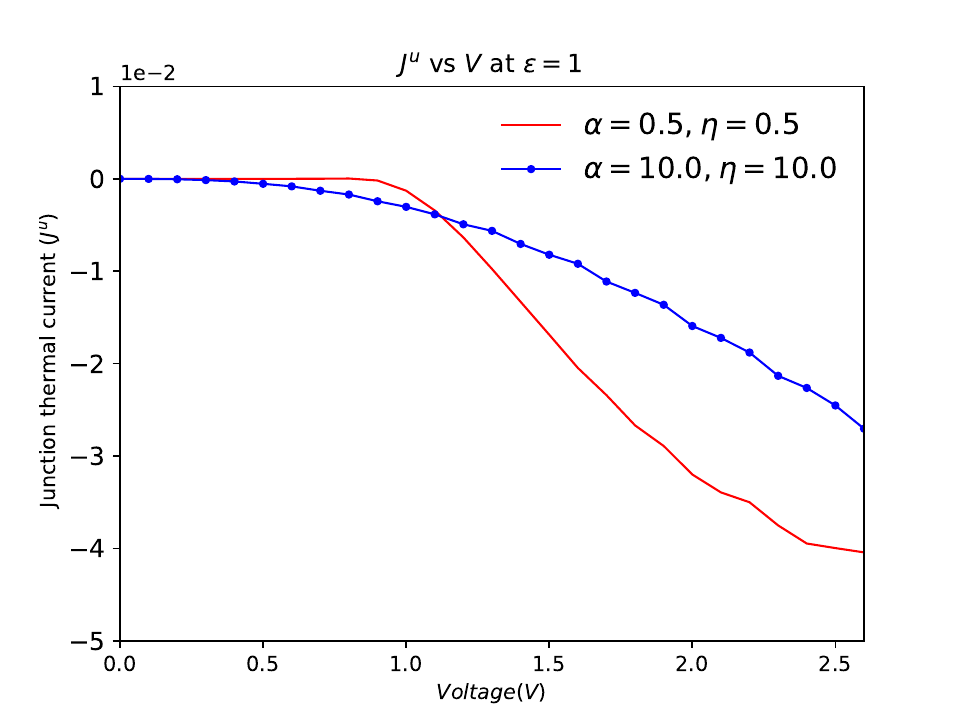}}
\caption{Plot of thermal currents ${J^u}$ vs $V$, $N=20, |{\epsilon}|=1.0, |\gamma_0|=0.5,\Delta=0.15,\gamma_{p} =1.0 , \gamma'_{p} =0.25 $}
\label{en_V_V1}
\end{figure}

\end{subsection}

\end{section}

\begin{section}{Conclusion}
\label{sec:conc}
We have studied the electrical, thermal, and thermoelectric transport through an open long-range Kitaev chain, whose left and right edges are connected to the left and right fermionic baths respectively. In different scenarios, baths are kept at different voltage/thermal biases, which drive electrical/thermal/thermoelectric current through the middle Kitaev chain. We observe that such transport through the LRK chain is greatly influenced by the presence of long-range interaction in the model. Long-range interaction influences subgap and bulk states in mainly two different ways, 
%(i) it enhances the overlap between the MBSs situated at the two distant edges of the middle chain and thereby paired them up onto a single massive Dirac fermion, 
(i) it prohibits complete gap closure at the topological phase transition point ($\epsilon=2 \gamma_0$ ) by giving mass to the subgap edge states (massive Dirac fermions), and (ii) it causes the delocalized bulk states to bunch up just above and below the band gap as compared to that of its short-range counterpart. We propose that these two distinct features of the LRK chain can be probed through current vs voltage and current vs $\Delta T $ characteristics. Firstly, at the TPT point, both electric currents ($J^e$) and thermal currents ($J^u$) in LRK chain remain almost zero in the initial part of the above-mentioned characteristic curves. However, as voltage or temperature biases become sufficiently large enough to excite the quasiparticles to overcome the energy bandgap which is directly proportional to the mass of the subgap state, both $J^e$ and $J^u$ start increasing with a faster rate as compared to that of SRK chain. This is an evidence of the absence of gap closure in the LRK chain at topological phase transition point which can be probed experimentally. This faster rate of increase of currents is a direct consequence of the clustering of bulk states just above the energy gap in the case of LRK chain.

%We find that the energy spectrum is greatly influenced by the  presence of long range interaction in LRK.  We observe that the modified energy spectrum of LRK chain significantly changes the electric/thermal/thermoelectrical current vs voltage/temp characteristics especially at TPT point which is absent in short-range limit.

On the other hand, this characteristics may be treated as a testing tool to understand the onset of long-range interaction in the LRK chain through transport measurement. Thus we provide an alternative method to probe long-range interaction by electric/thermal current measurements in addition to other conventional methods described in Ref. \cite{French_RMP}. Nature of bunching and delocalization of bulk states can also be probed by studying the slopes of $J^u$ vs $\Delta T$ and/or $J^e$ vs $V$ characteristic at TPT ($\epsilon=2 \gamma_0$ ) as the current increases at a faster rate in the case of long-range Kitaev chain compared to short-range Kitaev chain. To the best of our knowledge, this novel method of using electric/thermal/thermoelectric currents to investigate long-range interaction and its ramifications on the energy spectrum, has not been studied yet. Initial zero current region in the $J^e$ vs $V$ (or $J^u$ vs $\Delta T$ ) characteristic at TPT in LRK chain may provide a good estimation of mass of Dirac edge modes. 
 
Overall, the quantum transport through a long-range Kitaev chain features some unique properties that are absent in the short-range limit.  We expect these novel features which are important from the perspective of decoherence-free quantum computing using topologically protected modes, can be detected experimentally.

\end{section}

\begin*{\it Acknowledgments.}
NB acknowledges funding from DST-FIST programme.
\end*

\appendix
\setcounter{figure}{0}
\renewcommand\thefigure{A\arabic{figure}}
\section{Overview of LEGF Method}
\label{App1}
Quantum Langevin equations \& Green’s function method (LEGF) is a well known method to study non-equilibrium quantum transport in open-quantum system formulation within the Heisenberg representation of quantum mechanics. This method has been discussed thoroughly in Ref. \cite{DharSenPRB2006,RoyPRB2012,BondyopadhayaJSP2022,BhatDharPRB2020}. However, for the sake of completeness, we briefly discuss the essential parts of this method. First, we introduce a convenient basis ${\bf a}\equiv(a_1, a_2, \dots, a_{2N-1}, a_{2N})^T=(c_1, c^{\dg}_1, \dots, c_{N}, c^{\dg}_{N})^T$ to write the quadratic Kitaev chain Hamiltonian (\ref{HLRK}) as $\mathcal{H}=\frac{1}{2}{\bf a}^{\dg}\mathcal{H}^{\rm K}{\bf a}=\frac{1}{2}\sum_{l,m}\mathcal{H}_{lm}^{\rm K}a^{\dg}_la_m$  in terms of the matrix $\mathcal{H}_{lm}^{\rm K}$. 
%  Thus, $a_{2l}=a^{\dg}_{2l-1}$. Clearly the index $l$ in $a_l$ (or $a_l^\dagger$) does not represent actual physical site of the wire.  For a given $l$, one can define a map to the physical site $l'$ of spinless fermions as: $l'=(l+1)/2$ for odd values of $l$, and $l'=(l/2)$ for even values of $l$. 
We consider the left ($L$) and right ($R$) baths are in thermal equilibrium at temperatures and chemical potentials are given by ($T_L$, $\mu_L$) and ($T_R$, $\mu_R$), respectively, before we connect them through the middle Kitaev wire/chain at time $t_0$. Isolated semi-infinite bath Hamiltonians (\ref{HBath}) can also be written in a similar type of basis as  $H_{M}^p=\frac{1}{2}\sum_{\alpha,\beta}\, \mathcal{H}_{\alpha \beta}^{p}\, (a^p_\alpha)^{\dg}a^p_\beta$  with $\alpha,\beta=1,2,\dots,\infty$ ($p=L,R$). In a similar manner, tunnel coupling Hamiltonian $\mathcal{H}_T^p$ (\ref{HTun}) can also be written in this new basis. 

At time $t=t_0$, we connect both the baths to the Kitaev chain through tunnel couplings to construct N-TS-N hybrid device. Here, we are interested in the steady state properties of this chain. For $t > t_0$, the Heisenberg equations of motion for the chain and baths variables read
\bea
\dot{a}_l &=& -i \sum_{m=1}^{2N} \mathcal{H}^{\rm K}_{lm} a_m -i\gamma_{L}' (a_2^L \, \delta_{l,2}-a_1^L \, \delta_{l,1})-i\gamma_{R}'( a_2^R \, \delta_{l,2N}-a_1^R \, \delta_{l,2N-1}) \, ,
\label{eomwire}
\eea
for $l=1,\dots, 2N$, and 
\bea
 \dot{a}^L_{\alpha}  = -i \sum_{\beta} \mathcal{H}^{\rm L}_{\alpha \beta} \, a_{\beta}^{L} +i\gamma_{\rm L}'( a_1 \, \delta_{\alpha,1}-a_2 \, \delta_{\alpha,2}) \, , \nn \\
\label{eombathl}
 \eea
 for $\alpha = 1, \dots, \infty $, and
 \bea
 \dot{a}^R_{\alpha'} = -i \sum_{\beta'} \mathcal{H}^{\rm R}_{\alpha' \beta' }\, a_{\beta '}^R
 +i\gamma_{\rm R}' ( a_{2N-1} \, \delta_{\alpha' ,1}-a_{2N} \, \delta_{\alpha',2}  )\, , \nn \\
 \label{eombathr}
 \eea
for $\alpha' =1 \dots, \infty$. The Eqs.~\ref{eombathl}, \ref{eombathr} are coupled, inhomogeneous, first-order differential equations that can be formally solved for the boundary bath operators by using the retarded Green's function. Substituting these solutions for the bath operators into Eq.~\ref{eomwire}, one can rewrite Eq.~\ref{eomwire} in a form of generalized quantum Langevin equation \cite{BondyopadhayaJSP2022}. The quantum Langevin equations of Kitaev chain variables can be solved in the frequency domain using the Fourier transformation. To this end, we first consider the limit $t_0 \rightarrow -\infty$. The Fourier transform of the chain variables are defined as $\tilde{a}_l (\omega)=\frac{1}{2\pi}\int_{-\infty}^{\infty} dt \, a_l(t)\, e^{i \omega t}$. We get the following steady-state solutions for $\tilde{a}_l(\omega)$ after taking Fourier transform of the quantum Langevin equations of the Kitaev chain variables, $a_l(t)$:
\bea
\tilde{a}_l(\omega)&=&\sum_{m=1}^{2N} \tilde{G}^+_{l,m}(\omega)\left( \sum_{k=1,2} \tilde{\eta}_{k}^{\rm L} (\omega)\, \delta_{m,k} +\sum_{k=1,2} \tilde{\eta}_{k}^{\rm R} (\omega)\, \delta_{m,2N+k-2} \right) ,\nn \\
\label{a1}
\eea 
%\end{widetext}
where $l=1,\dots,2N $ and $\tilde{G}^+(\omega)$ is the retarded Green's function of the full system. Here, $\tilde{\eta}^{\rm L}_{1,2}(\omega) $ and $\tilde{\eta}^{\rm R}_{1,2}(\omega) $ are the noise terms arising in the process of integrating out the variables of $L$ and $R$ bath, respectively. These noise terms keep track of the non-equilibrium boundary conditions across the middle wire, which we impose in the beginning through the initial equilibrium correlators for baths $\langle c^{p \dg}_\beta (t_0)c^p_{\beta '} (t_0) \rangle$ (\ref{ini}) for $p=L, R$. Noise-noise correlators in frequency domain are given by \cite{DharSenPRB2006,BhatDharPRB2020,BondyopadhayaJSP2022}
\bea
\langle \tilde{\eta}^{\rm p \dg}_{1}(\omega)  \tilde{\eta}^{\rm p}_{1}(\omega ') \rangle &=& -(\gamma_p'^2/\pi) {\rm Im}[\tilde{G}^{p+}_{1,1}(\omega)] f(\omega, \mu_p,T_p) \delta(\omega- \omega') \, ,\nn \\
\langle \tilde{\eta}^{\rm p \dg}_{2}(\omega)  \tilde{\eta}^{\rm p}_{2}(\omega ') \rangle &=& -(\gamma_p'^2/\pi) {\rm Im}[\tilde{G}^{p+}_{2,2}(\omega)] f(\omega,-\mu_p,T_p) \delta(\omega- \omega')\, , \nn \\
\langle \tilde{\eta}^{\rm p \dg}_{1}(\omega)  \tilde{\eta}^{\rm p}_{2}(\omega ') \rangle & =& \langle \tilde{\eta}^{\rm p \dg}_{2}(\omega)  \tilde{\eta}^{\rm p}_{1}(\omega ') \rangle =0 \, . \nn
\eea
Here, $\tilde{G}^{p +}_{l,m}(\omega)$ is the retarded Green's function of isolated bath ($p ={L,R}$). $\tilde{G}^{\rm L +}_{1,1}(\omega)$ and  $\tilde{G}^{\rm L +}_{2,2}(\omega)$
correspond to the first site of the left bath which is connected to the first site of the Kitaev wire, whereas $\tilde{G}^{\rm R +}_{1,1}(\omega)$ and  $\tilde{G}^{\rm R +}_{2,2}(\omega)$ correspond to the first site of the right bath which is connected to the $N$-th (last) site of the Kitaev chain. It can be shown that, within the bandwidth of the bath ($ |\omega| < 2 \gamma_p$), retarded Green's function for isolated baths are given by \cite{RoyPRB2012},
\beq
\tilde{G}^{p +}_{1,1}(\omega)=\tilde{G}^{p +}_{2,2}(\omega)=\frac{1}{\gamma_p} \left[ \frac{\omega}{2 \gamma_p}-i\left( 1- \frac{\omega^2}{4 \gamma_p^2} \right)^{1/2} \right], \nn
\eeq
where $p=L, R$.
%The detailed definitions and expressions of $\tilde{G}^{\alpha +}_{l,m}(\omega)$ are given in Appendix~\ref{App2}.

The retarded Green's function $\tilde{G}^+(\omega)$ of the full system in the Fourier domain is defined as
\bea
\tilde{G}^+(\omega)&=&{(\omega \one_{2N} - \mathcal{H}- \tilde{\Sigma}_{\rm L}^+(\omega) -\tilde{\Sigma}^+_{\rm R}(\omega))^{-1}} \nn \\
&=&{(\omega \one_{2N} - \tilde{\mathcal{H}})}^{-1}\,,
\label{FG0}
 \eea
where $\tilde{\Sigma}^+_{\rm L,R}$ are the self-energy corrections to the Kitaev chain Hamiltonian originated from its interactions with the respective baths. The effective Hamiltonian matrix of the Kitaev chain is given by $\tilde{\mathcal{H}}=\mathcal{H}+ \tilde{\Sigma}^+_{\rm L}(\omega) +\tilde{\Sigma}^+_{\rm R}(\omega)$.
The components of the self-energy terms $\tilde{\Sigma}^+_{\rm L,R}$ are as following:
\bea
[\tilde{\Sigma}_{\rm L}^+(\omega)]_{lm} &=& {\gamma_{\rm L}'}^2 \sum_{k=1}^{2} {\tilde{G}^{\rm  L +}_{k,k}(\omega)}\, \delta_{l,3-k}\, \delta_{l,m}\, ,\nn \\
{[\tilde{\Sigma}^+_{\rm R}(\omega)]}_{lm} &=& {\gamma_{\rm R}'}^2 \sum_{k=1}^{2}{\tilde{G}^{\rm R+}_{k,k}(\omega)}\,\delta_{l,2N-2+k}\, \delta_{l,m}\, , \nn \\
\eea
where $l,m= 1,\dots,2N $. Since $\tilde{\mathcal{H}}$ is a block diagonal matrix, numerical values of $\tilde{G}^+(\omega)$ can be calculated by inverting $(\omega \one_{2N} -\tilde{\mathcal{H}})$.

We further write steady-state solutions for some of the bath variables $\tilde{a}_l^{L/R}(\omega)$ defined at the edges of the baths. For example, these $\tilde{a}_{1}^L(\omega )$ and $\tilde{a}_{2}^L(\omega )$ for the left bath read 
\bea
\tilde{a}_{1}^L(\omega )\gamma_{\rm L}' &=&  -\tilde{\eta}^{\rm L}_1(\omega)-
\sum_{m=1}^{2}{[\tilde{\Sigma}^+_{\rm L}(\omega)]}_{1,m} \, \tilde{a}_{m}(\omega),\nn \\
 \tilde{a}_{2}^L(\omega )\gamma_{\rm L}' &=&  \tilde{\eta}^{\rm L}_2(\omega)+
\sum_{m=1}^{2}{[\tilde{\Sigma}^+_{\rm L}(\omega)]}_{2,m} \, \tilde{a}_{m}(\omega) \, .\nn
\label{bathvariable}
\eea
These boundary variables of the baths are useful in evaluating the currents through the N-TS-N device. First taking the Fourier transformation of the currents expressions (\ref{elec_c},\ref{energy_c}), then substituting chain variables ($\tilde{a}_l(\omega)$) and bath variables ($\tilde{a}_\alpha^{L/R}(\omega)$) in the Fourier transformed expressions, and finally using noise-noise correlations for baths, one can calculate the steady-state currents ($J^e_{L/R}$, $J^u_{L/R}$) in N-TS-N system \cite{BondyopadhayaJSP2022}.

\bibliography{bibliographyL}
\end{document}